\def\aj{{AJ}}

\def\apj{{ApJ}}

\def\mnras{{MNRAS}}
\def\nat{{Nature}}
\def\pasp{{PASP}}

\documentclass[cup5b]{caps}
\usepackage{graphicx}
\usepackage{amssymb}
\usepackage{ociwsymp2}   

\HeadText{A. C. S.  Readhead and T. J.  Pearson}

\begin{document}

\pagenumbering{arabic}

\author[]{ANTHONY C. S. READHEAD and TIMOTHY J. PEARSON
\\California Institute of Technology}

\chapter{Interferometric Observations of the\\ Cosmic Microwave Background
Radiation}

\begin{abstract}

Radio interferometers are well suited to studies of both total intensity and
polarized intensity fluctuations of the cosmic microwave background radiation, and they
have  been used successfully in measurements of both the primary and secondary anisotropy. 
Recent observations with the Cosmic Background Imager operating in the
Chilean Andes, the Degree Angular Scale Interferometer operating at the
South Pole, and the Very Small Array operating in Tenerife have probed the primary anisotropy over a wide range of
angular scales.  The
advantages of interferometers for microwave background observations of
both total intensity and polarized radiation are discussed, and the
cosmological results from these three instruments are presented. The
results show that, subject to a reasonable value for the Hubble constant, which is degenerate with the geometry in
closed models,  the geometry of the Universe is flat to high
precision ($\sim 5\%$) and the primordial fluctuation spectrum is very
close to the scale-invariant Harrison-Zel'dovich spectrum. Both of these findings
are concordant with inflationary predictions. The results also show that the
baryonic matter content is consistent with that found from primordial
nucleosynthesis, while the cold dark matter component can account for no
more than
$\sim 40 \%$ of the energy density of the Universe.  It is a
requirement of these observations, therefore, that  $\sim 60\%$ of the
energy content of the Universe is not related to matter, either baryonic
or nonbaryonic. This {\it dark energy} component of the energy density
is attributed to a nonzero cosmological constant.

\end{abstract}

\section{Introduction}

Interferometers are playing a key role in the determination of
fundamental cosmological parameters through observations of the cosmic microwave
background (CMB).  Three such instruments  have been
constructed and deployed over the last few years --- the Cosmic Background
Imager (CBI), operating at  5080 m altitude in
the Chilean Andes (Padin et~al.\ 2001, 2002); the Degree Angular Scale Interferometer (DASI),
operating at 2800 m altitude at the South Pole (Leitch et~al.\ 2002a); and the Very Small Array (VSA)
operating at an altitude of 2400 m in Tenerife (Watson et~al.\ 2003), for which a prototype was the
Cambridge Anisotropy Telescope (CAT) (Scott et~al.\ 1996).  In this paper we review the characteristics of
interferometers that render them particularly well suited to CMB observations,
and we then discuss the
cosmological results from observations with these three instruments. 

\begin{figure*}[t]
\centering
\includegraphics[width=0.97\columnwidth,angle=0,clip]{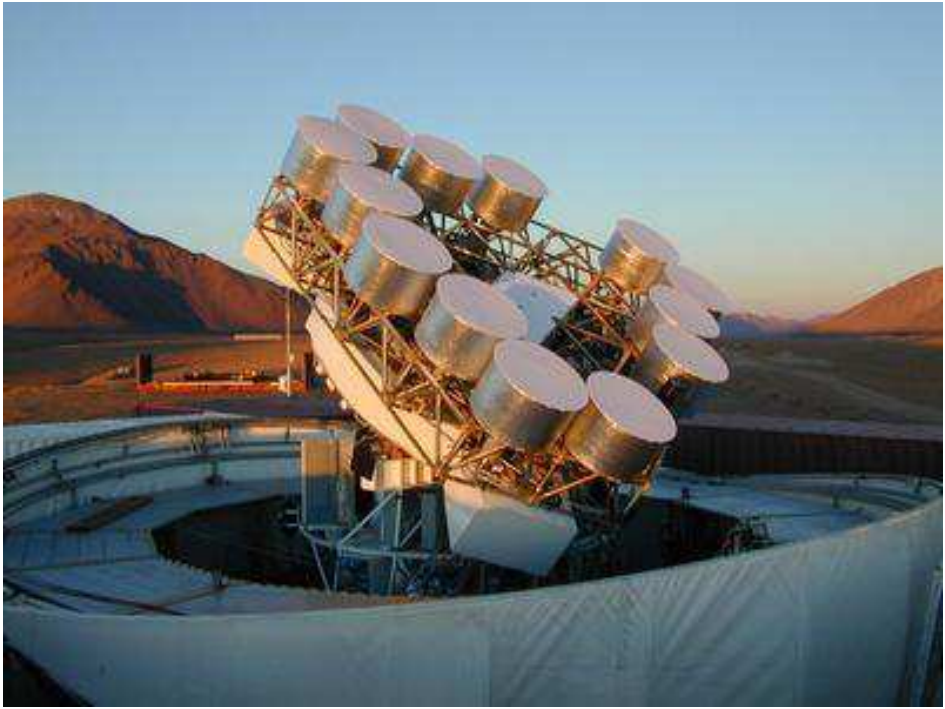}
\vskip 0pt \caption{
The Cosmic Background Imager (CBI), located at 5080 m altitude in the Chilean 
Andes, was the first instrument to operate permanently from this, the future 
ALMA site.  The CBI is a ``stand-alone'' operation with its own power plant 
and other observatory facilities.  Oxygenated working and living quarters, 
pioneered here, have proven essential to the project.
\label{fig:cbi}}
\end{figure*}

 It is well known that bolometers have also played a crucial role in the development of this field, and
the interested reader will find an account of the bolometer CMB results in the companion review in this volume by Lange.  While we do not
discuss the bolometer results here, they are in excellent agreement with the interferometer
results described below (de~Bernardis et~al.\ 2000; Hanany et~al.\ 2000; 
Lee et~al.\ 2001; Netterfield et~al.\ 2002), and
all of these results are in excellent agreement with the recently released 
{\it WMAP} results (Bennett et~al.\ 2003).

\begin{table*}[t]
\centering
\caption{Specifications of CMB Interferometers}
\begin{tabular}{c|ccc}
\hline \hline
     {Specification} & {${\rm CBI}$} & {${\rm DASI}$} & {${\rm VSA}$}\\
     \hline
     antennas & 13
       & 13 & 14\\
     center frequency  (GHz)& 31    & $31$ & 31\\
     channels & 10 & 10 & 1(tunable)\\
     bandwidth per channel (GHz) & 1    & 1    & 1.5 \\
    system temperature & $\sim 30$ K & $\sim 30$ K & $\sim 30$ K\\
     $l$ range & 300--3500& 150--850 & 150--1400 \\
     $l$ resolution & $\sim 140$  & $\sim 80$ & $\sim 80 $ \\
     location & Atacama Desert & South Pole & Tenerife \\
    altitude (m) & 5080 & 2800 & 2400 \\
\hline \hline
\end{tabular}
\label{tab:specs}
\end{table*}

\section{The CBI, the DASI, and the VSA}
The CBI, shown in Figure \ref{fig:cbi},  was the first of the new generation of CMB interferometers to be
brought into operation
(Padin et~al.\ 2001), in January 2000.  At that time the CBI began routine total intensity observations
as well as preliminary polarization observations using a single cross-polarized antenna. We describe
the CBI here in
some detail since both the DASI and the VSA share most of these characteristics, and we point out the
principal features in the VSA and the DASI in which they differ from the CBI. 
 All three instruments operate in the Ka waveguide
band (26 -- 40 GHz) and use low-noise amplifiers (LNAs) based on high electron
mobility transistors (HEMTs), which were designed by Marian  Pospieszalski of the
National Radio Astronomy Observatory (Pospieszalski et~al.\ 1994, 1995). The specifications of the three
instruments are given in Table \ref{tab:specs}.

As described by Padin et~al.\ (2002), the CBI has 13 90-cm diameter antennas mounted on  a
6-m platform. In addition to the usual altazimuth rotation
axes, the CBI platform can be rotated about the optical axis.  This third axis of rotation enables the array to
maintain a constant parallactic angle while tracking a celestial source; it also greatly
facilitates polarization observations and  calibration, and,
in addition, it provides a powerful method of discriminating between sky signals and instrumental
cross-talk. For each baseline both the real and imaginary channels are correlated in a complex correlator, and
there are 10 1-GHz bandwidth frequency channels spanning the range 26--36 GHz.  Thus, the CBI
comprises a total of 780 complex interferometers.  The fundamental data rate is 0.84~s.  For typical total intensity observations
the data are
averaged over 10 samples, so that on each interferometer the complex
visibility is measured every 8.4 s.

\begin{figure*}[t]
\centering
\includegraphics[width=1.00\columnwidth,angle=0,clip]{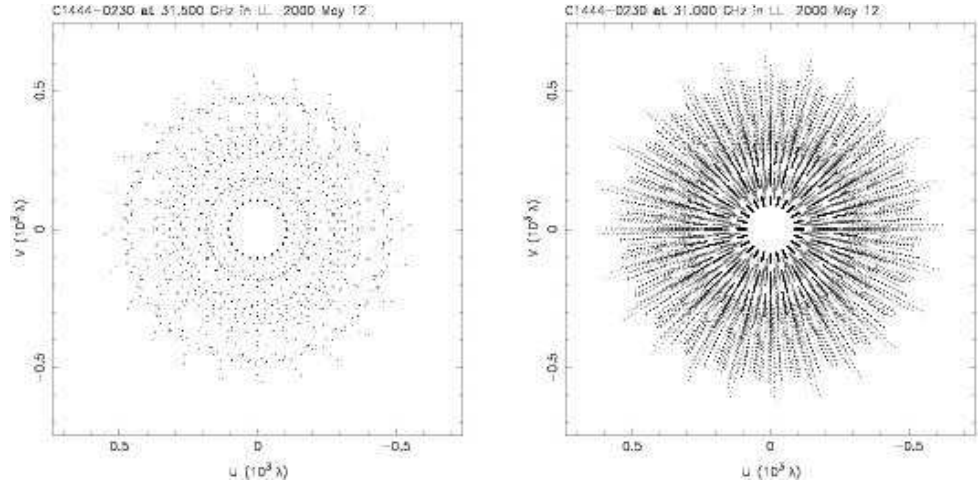}
\vskip 0pt \caption{
The aperture-plane, ({\it u,v}), coverage of the CBI for a 6-hour 
observation.  The left-hand panel shows the coverage for a single channel, and 
the right-hand panel shows the coverage for all 10 channels.
\label{fig:cbiuv}}
\end{figure*}

\begin{figure*}[b]
\centering
\includegraphics[width=1.00\columnwidth,angle=0,clip]{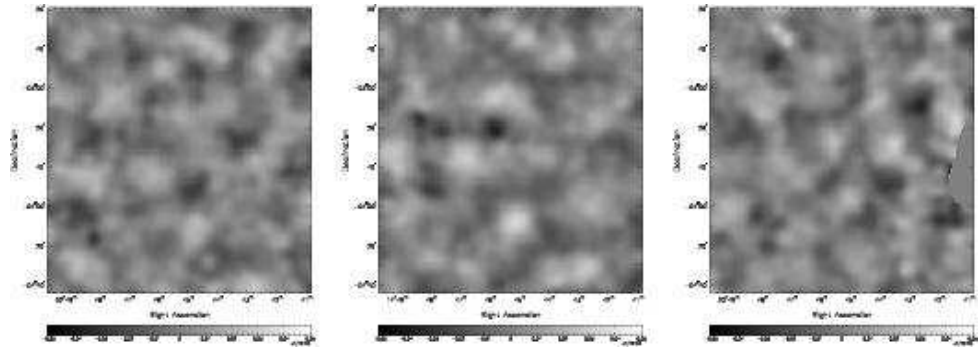}
\vskip 0pt \caption{
The CBI has made mosaic observations of three regions separated by 6 hr in
Right Ascension.  Mosaic images from the first year of observations, which
cover an area of $\sim 2^{\circ} \times 2^{\circ}$, are shown above. In these
images the seeds that gave rise to clusters of galaxies are seen for the
first time.
\label{fig:mosaics}}
\end{figure*}

The aperture-plane, or ({\it u,v}), coverage of the CBI is shown in Figure \ref{fig:cbiuv}.  In this
figure the ({\it u,v}) coverage corresponding to a single baseline should be the convolution of two apertures centered on the ({\it u,v})  point
corresponding to the baseline length and orientation.  For clarity we have shrunk the aperture convolution to a single point
 to make the distribution of the coverage clearer.  There is, of course, much overlap in the coverage
when the full diameters are used, since the antenna diameter of 90 cm is close to the size of the shortest baselines (100 cm).  Mosaic observations
of contiguous fields improve the resolution in multipole space. For example, in the CBI, mosaic observations have been used to improve the
multipole resolution from
$\delta l \approx 500$ to
$\delta  \approx 140$. CBI mosaic observations of three fields
are shown in Figure
\ref{fig:mosaics}.
 
\begin{figure*}[t]
\centering
\includegraphics[width=0.97\columnwidth,angle=0,clip]{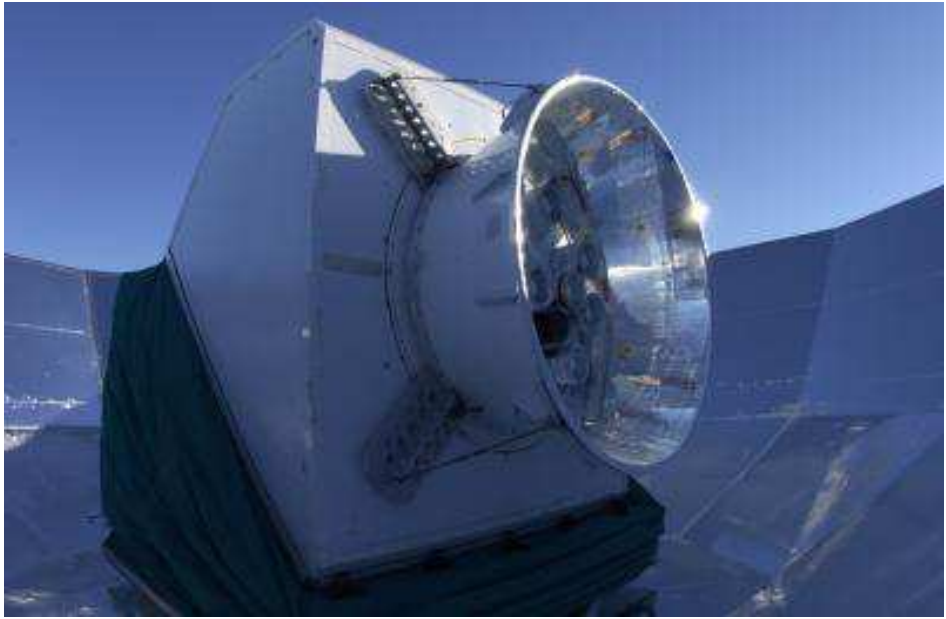}
\vskip 0pt \caption{
The Degree Angular Scale Interferometer (DASI), located at 2800 m at the South
Pole, was outfitted with a sunscreen for the polarization observations of 2001.
\label{fig:dasi}}
\end{figure*}

The DASI is shown in Figure \ref{fig:dasi}.  The DASI was designed to complement the CBI in multipole
coverage, and many of the detailed CBI designs, including those for the correlator, the receiver
control cards, and the channelizer, as well as the telescope control and data acquisition software, were
shared with the DASI team, who duplicated these components in the DASI.  For these reasons, the DASI is
in many respects a copy of the CBI, but $\sim$ 4 times smaller.  The DASI achromatic polarizers were
duplicated by the DASI team for use on the CBI in its upgraded phase for full polarization
observations, which commenced in October 2002. The DASI observed in total intensity mode during the
austral winter of 2000, and was then outfitted with achromatic polarizers and both a ground screen and
a sunscreen, shown in Figure \ref{fig:dasi},  for the austral winter of 2001 when it made polarization
observations of two fields, as described by Zaldarriaga (2003).

\vfil
\begin{figure*}[b]
\centering
\includegraphics[width=0.97\columnwidth,angle=0,clip]{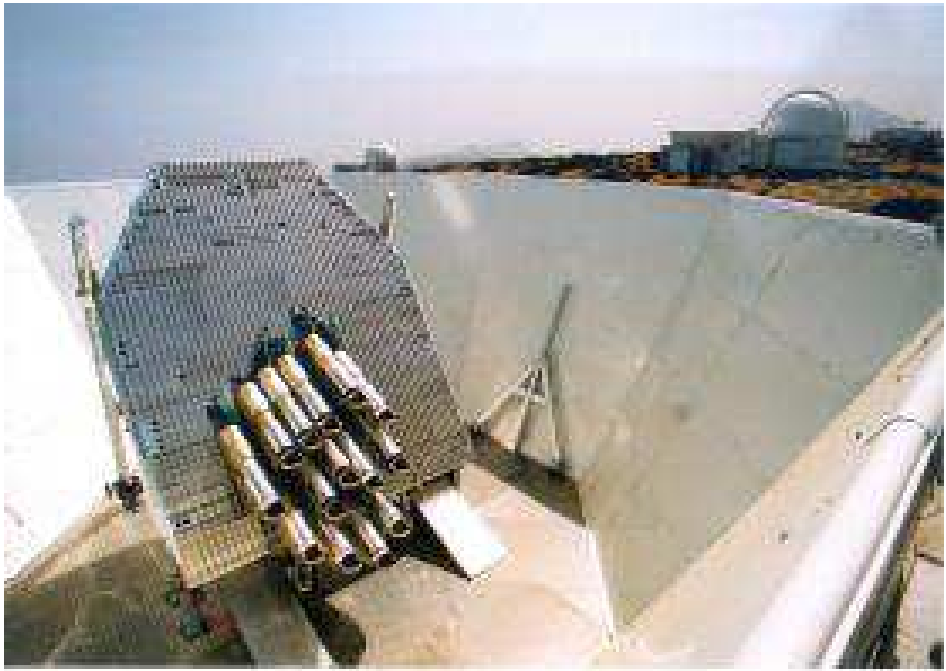}
\vskip 0pt \caption{
The Very Small Array (VSA) is sited at 2400 m in Tenerife.
\label{fig:vsa}}
\end{figure*}

\clearpage

The VSA (Figure \ref{fig:vsa}) and the DASI are similar in size, and originally covered similar multipole ranges
(and hence very similar angular scales), extending from $l\approx 150$
to $l \approx 900$ with a resolution in multipoles of $\delta l \approx 80$.
In 2001 the original VSA horns were replaced with a larger set that extended the VSA range up to $l
\approx 1400$.    
 
The CBI
covers the range $l \approx 300 $ to $l \approx 3500$ with a multipole
resolution in mosaicing mode  of $\delta l \approx 140$.  Apart from this primary difference in multipole
range and resolution, the other major difference is in the correlators --- the CBI and the
DASI each have 10 1-GHz channels, operating between 26 GHz and 36 GHz, while the VSA
has a single tunable 1.5-GHz channel, which operates over the same frequency range.

\section{The Sky Brightness Distribution of the CMB}
We denote the CMB temperature in direction ${\bf \hat n}$ (angular coordinates
$\theta,\phi$) by $T({\bf \hat n})$, and hence the variation in temperature about
the
mean $\langle T({\bf \hat n}) \rangle$ by

\begin{equation}
 \Delta ({\bf \hat n})= {T({\bf \hat n}) \over {\langle T({\bf \hat n})
\rangle}} -1, 
\end{equation}
which may be expanded into spherical harmonics
\begin{equation}
 \Delta ({\bf \hat n})= \sum_{l=0}^{\infty} \sum_{m=-l}^l
a_{lm}Y_{lm}({\bf \hat n}), 
\end{equation}
where $a_{lm}$ are the multipole moments
\begin{equation}
a_{lm}=\int Y_{lm}^*({\bf \hat n}) \Delta({\bf \hat n})d\Omega \quad .
\end{equation}
If the CMB is isotropic it must be rotationally invariant so that the fluctuations
can be expressed in terms of a one-dimensional angular spectrum, $C_l \equiv \langle a_{lm}^2 \rangle$, and the
variance of $\Delta({\bf \hat n})$ can be expressed in terms of this angular
power spectrum:
\begin{equation} 
\langle \Delta({\bf \hat n})^2\rangle=\sum_{l=0}^{\infty}{2l+1 \over {4
\pi}}C_l \quad .
\end{equation}
For large $l$, 
\begin{equation} 
\langle \Delta({\bf \hat n})^2\rangle \approx \int {l(2l+1) \over {4
\pi}}C_l\; d\, {\rm ln}\,l, 
\end{equation}
showing that the contribution to the variance from a logarithmic interval of $l$
is proportional to $l(2l+1)C_l/4\pi$. It is therefore often convenient to plot this quantity
versus log$\,l$, so this convention is adopted by many authors in displaying angular spectra of the CMB.  It should be borne in mind,
however, that the signal is
$C_l$, so that, in plotting $l(2l+1)C_l/4\pi$ we are artificially boosting the
apparent variance at high $l$ by a factor $\propto l^2$.  {\it Observations at high $l$
therefore require far greater sensitivity than is immediately apparent from this
conventional way of plotting the angular power spectrum.}
 
The angular correlation function is defined as
\begin{equation}
C(\theta)=C({\bf \hat n_1},{\bf \hat n_2})=\langle  \Delta({\bf \hat
n_1}) \Delta({\bf
\hat n_2})\rangle, 
\end{equation}
where by isotropy and homogeneity $C$ is a function only of $\theta$, and
$\cos\theta=
{\bf \hat n_1}\cdot {\bf \hat n_2}$.
It is easy to show that
\begin{equation}
C(\theta)={1 \over {4\pi}}\sum_{l=0}^{\infty}(2l+1)C_lP_l(\cos\theta), 
\end{equation}
where $P_l(\cos\theta)$ is the Legendre polynomial.  The inverse result is
\begin{equation}
C_l=2 \pi \int_{-1}^1 C(\theta)P_l(\cos\theta)d\,\cos\,\theta \quad .
\end{equation}
These two results are the analogs on the sphere of the theorem relating the
covariance function to the Fourier transform of the power spectrum.

In interferometric observations of the CMB the fields of view, even for
mosaic observations, are small enough that the small-angle
 approximation may be used.  That is, we can  assume that
\begin{equation}\theta \ll 1 \; \;{\rm and} \;\; l \gg 1 .
\end{equation}
  In this case
\begin{equation}
P_l(\cos \theta) \approx J_0\{(l+{1/ 2})\;\theta\},  
\end{equation}
so that
\begin{eqnarray}
C_l & \approx&  2 \pi \int_{-1}^1 C(\theta)J_0\{(l+{1/ 2})\;\theta\} 
d\,\cos\,\theta \nonumber \\
& \approx & 2 \pi \int_{0}^{\infty} C(\theta)J_0\{(l+{1/ 2})\;\theta\}\,\theta\, 
d\,\theta \quad .
\end{eqnarray}

If we set  $l+1/2= 2 \pi v$ then this is a Hankel transform, which is simply the
two-dimensional Fourier transform for a circularly symmetric function; i.e., as
expected, the angular correlation
function and the power spectrum form a Fourier transform pair in the
small-angle approximation.
\vskip 6pt \noindent

In practice we observe the sky with instruments of finite size, so that what we
observe is the convolution of the sky brightness distribution with the instrument
response function, corresponding to a product in the Fourier transform (spectral) domain:
\begin{equation}
C(\theta)={1 \over {4\pi}}\sum_{l=0}^{\infty}(2l+1)C_l W_l(\cos\theta), 
\end{equation}
where $W_l(\cos \theta)$ is called the window function of the instrument.

\section{Radio Interferometric Observations of the CMB}

We do not review here the basic theory of radio interferometry, but refer
the reader to the comprehensive text of Thompson, Moran, \& Swenson (2001).  
Useful discussions of the application of standard
radio interferometric techniques to observations of the CMB have  been given
by Hobson, Lasenby, \& Jones (1995), White et~al.\ (1999), Hobson \& Maisinger 
(2002), and Myers et~al.\ (2003).

\begin{figure*}[t]
\centering
\includegraphics[width=0.40\columnwidth,angle=0,clip]{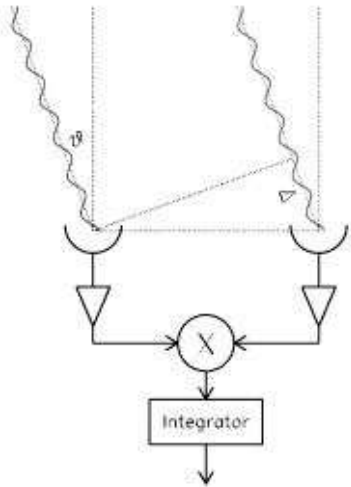}
\vskip 0pt \caption{
The Multiplying Interferometer subtracts off the constant term in the 
field under observation, as was first demonstrated by Martin Ryle in 1952 
(see text).
\label{fig:interferometer}}
\end{figure*}

\section{Interferometer Characteristics}

The following properties of interferometers make them particularly well suited to
observations of the CMB:
\begin{itemize}
 \item  Automatic subtraction of the mean signal (to high precision)
\item  Precise knowledge of the beamshape is easy to obtain (and is not as important as it is in single-dish observations)
\item  Direct measurements of visibilities (which are very nearly the desired Fourier components of
the sky brightness distribution)
\item  Precision radiometry (through observations of planets, supernova remnants, quasars, and radio galaxies)
\item  Precision polarimetry (and, in the case of the CBI, making use of the third axis of rotation about the optical axis)
\item  Repeated baselines enable a wide variety of instrumental crosschecks
\end{itemize}
We discuss each of these interferometer properties separately below, illustrating some of them with examples
from the CBI.  It should be clear that similar advantages apply, in most cases, to the DASI and the VSA. 
 
\subsection{Automatic Mean Subtraction}
A multiplying
interferometer (Fig. \ref{fig:interferometer}) has the advantage over adding interferometers and other
total-power detection systems in that the mean signal is subtracted automatically to high precision 
(Ryle 1952). In the adding interferometer the voltages from the two antennas are added and then squared in a square-law detector, so that the power
output from the square-law detector is proportional to $(V_1+V_2)^2$.  Ryle introduced the phase-switched interferometer
in which the voltages are alternately summed and differenced by introducing a $\pi/2$ phase offset in one of the signals, and then detected by a phase-synchronous
detector, which is in phase with the phase switch.  In this system the output power is the time average of $(V_1+V_2)^2-(V_1-V_2)^2$, and so it is proportional the
the product of the voltages from the two antennas.  The power output in this system is thus independent of the mean level of the signal, $V_1^2+V_2^2$,
 and it  measures the correlation between the two signals, $\langle V_1V_2 \rangle$.  Modern interferometers, such as those discussed here, accomplish the same
effect by using a correlator.  Since,  in a multiplying or phase-switching interferometer, the mean signal does not appear, and only the 
spatially varying signal appears,   this  eliminates many sources of spurious systematic errors.
This approach is particularly advantageous in CMB observations, in which the spatial
fluctuations in temperature are over
$ 10^5$ times 
smaller than the mean signal of 2.725 K (Mather et~al.\ 1999).  For this
reason a number of potential sources of systematic error are reduced to negligible
levels by interferometry. 

\begin{figure*}[t]
\centering
\includegraphics[width=1.00\columnwidth,angle=0,clip]{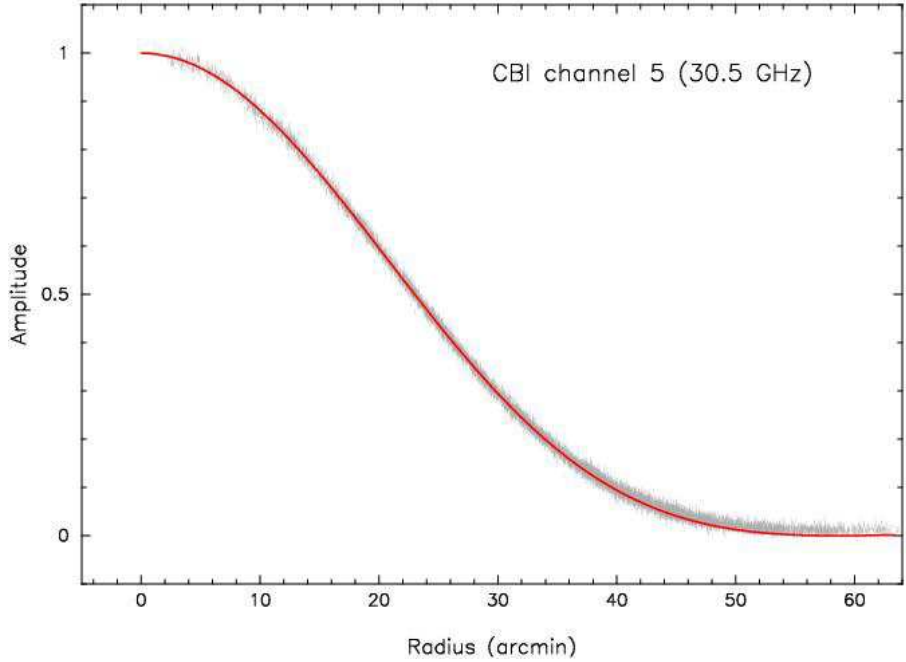}
\vskip 0pt \caption{
Radial profile of the CBI primary beam in one of the 10 frequency channels;
data from all 78 baselines are superimposed. The observations were of Tau~A
and are shown by the error bars.   The curve shows the profile computed by
taking the square of the Fourier transform of the aperture illumination
pattern, assumed to be circularly symmetric. These observations show that the
13 antennas have very similar beams (see Pearson et~al.\ 2003).
\label{fig:beam}}
\end{figure*}

\subsection{Precise Knowledge of Beamshape}
The resolution of an interferometer is set by the baseline length between the antenna pair, rather than by the
primary beam of the individual antennas. Thus, in interferometric observations precise knowledge of
the primary beamshape is not required in order to measure the power spectrum on small angular scales.  Of
course, precise knowledge of the primary  beamshape is needed in computing the covariance function of the
instrument, but this can be determined to the required accuracy of a few percent by measurements of the beamshape
on bright unresolved radio sources.  The profile of the mean primary beamshape for the CBI is shown in
Figure \ref{fig:beam}.  

\subsection{Direct Measurements of Visibilities}

  An interferometer measures visibility, which is the Fourier
transform of the product of the primary antenna beam, $A({\bf x})$, with the sky
brightness, $I({\bf x})$.  The visibilities,
$V({\bf u})=V(u,v)$,  measured by an interferometer are related to the sky
brightness by
\begin{equation}
V({\bf u})=\int_{-\infty}^{\infty}\int_{-\infty}^{\infty}A({\bf x})I({\bf x})\,
e^{-2\pi i{\bf u}\cdot{\bf x}}d{\bf x} \quad .
\end{equation}

\begin{figure*}[t]
\centering
\includegraphics[width=1.00\columnwidth,angle=0,clip]{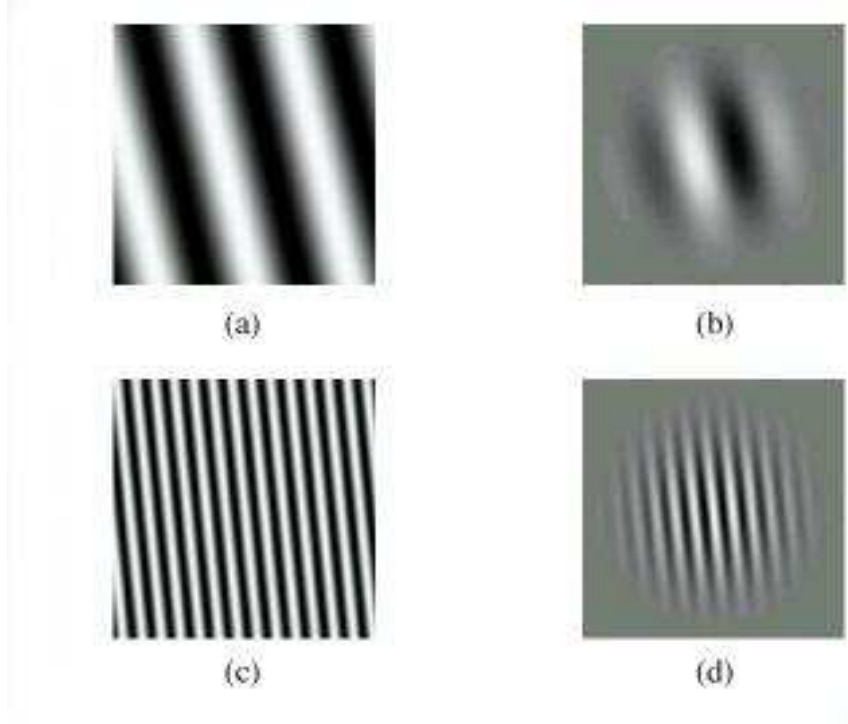}
\vskip 0pt \caption{
Panels (a) and (c) illustrate two multipole components of the sky brightness 
distribution over a $1.5^{\circ} \times 1.5^{\circ}$ field of view, typical 
for the CBI.  An interferometer measures directly these components multiplied 
by the primary beam, which is fixed by the antenna size and illumination.  The 
product of the primary beam with the multipole component measured by an 
interferometer, which is set by the baseline length, is shown for the two 
multipole components above in panels (b) and (d).  In the case of the CBI, 
(a) and (b) represent the one-meter baselines, while (c) and (d) represent the 
five-meter baselines.
\label{fig:interferometry}}
\end{figure*}

The visibility is thus the convolution of the Fourier transform of the sky
brightness,
$\tilde I({\bf u}) $ (i.e., the angular spectrum of the sky brightness
distribution), and of the primary beam,
$\tilde A({\bf u})$,
\begin{equation}
V({\bf u})=\tilde I({\bf u}) * \tilde A({\bf u}) \quad .
\end{equation}

An interferometer therefore measures directly the angular spectrum of the
sky brightness distribution, which is the desired result, convolved with the
Fourier transform of the primary beam.  An example of a single Fourier component on the sky and of this component
multiplied by the primary beam is shown in Figure \ref{fig:interferometry}.

The matrix of the covariance between all the visibility measurements is the
observed {\it covariance matrix}, $C_{ij} = \langle V_iV_j^*\rangle$, and this is
made up of two parts, 
\begin{equation}
{\bf C=M+N},
\end{equation}
where ${\bf M}$ is the sky covariance matrix and ${\bf N}$ is the noise covariance
matrix.  Hobson et~al.\ (1995) have shown that the sky covariance matrix
is
\begin{eqnarray}
M_{jk}&=&\langle V({\bf u}_j, \nu_j)V^*({\bf u}_k, \nu_k)\rangle \nonumber \\
      &=&\int \int d^2{\bf v} \tilde A({\bf u}_j - {\bf v},\nu_j) 
\tilde A({\bf u}_k - {\bf v},\nu_k)S({\bf v},\nu_j,\nu_k), 
\end{eqnarray}
where $S({\bf v},\nu_j,\nu_k)$ is a generalized power spectrum of the intensity
fluctuations. Hence, converting to temperature fluctuations, we find that for CMB fluctuations,
\begin{eqnarray}
M_{jk}&=&\langle V({\bf u}_j, \nu_j)V^*({\bf u}_k, \nu_k)\rangle \nonumber \\
      &=&\left( {2 \nu^2 k_B T_0 g(\nu) \over c^2 }\right)^2 \int \int 
         d^2{\bf v} \tilde A({\bf u}_j - {\bf v},\nu_j) \tilde A({\bf u}_k - 
         {\bf v},\nu_k)C(v) \nonumber \\
      &=&\left( {2 \nu^2 k_B T_0 g(\nu) \over c^2 }\right)^2 \int_0^{\infty} 
         W_{jk}(v) C(v) v dv, 
\end{eqnarray}
where $g(\nu)=x^2e^x/(e^x-1)^2$,  $x=h\nu /k_B T_0$, and $C(v)=C_l$ for
$\sqrt{l(l+1)} \approx l+{1 \over 2} = 2 \pi v$, and 
\begin{equation}
W_{jk}(v) = \int_0^{2 \pi} \tilde A({\bf u}_j - {\bf v},\nu_j) 
\tilde A({\bf u}_k - {\bf v},\nu_k)d \theta_v
\end{equation} 
is the {\it window function}.

\subsection{Precision Radiometry}
A great advantage of interferometric measurements is precision radiometry.  Provided that there are
sufficiently bright unresolved and nonvarying radio sources, the calibration of interferometers is
straightforward.  In the case of the CBI, for example, the primary calibrators were Jupiter and Tau A, and
secondary calibrators were Mars, Saturn, and 3C~274.  Of these only Tau A is significantly resolved on the
CBI, but it is easily modeled with an elliptical Gaussian brightness profile.  The internal
calibration consistency on the CBI is considerably better than 1\%, and the flux density scale, originally set by our own
absolute calibrations at the Owens Valley Radio Observatory with an uncertainty of 3.3\% (Mason et~al.\ 1999), 
has now been improved by comparison with the {\it WMAP} temperature measurement of Jupiter (Page et~al.\ 2003) to an uncertainty of 1.3\% (Readhead et~al.
2003).

\subsection{Precision Polarimetry}
The instrumental polarization of radio interferometers is much easier to calibrate than single-dish instruments, because there is an elegant way
of distinguishing between the polarization of the instrument and that of the source (Conway \& Kronberg 1969). 
The instrumental polarization can be expressed in terms of {\it leakage factors}.  In the CBI each antenna observes either left- or
right-circular polarization.  The voltage output from a left-circularly polarized antenna, $p$, may be written
\begin{equation}
V_L(p)=E_L(p) + E_R(p)\epsilon_p e^{j \phi_p}, 
\end{equation}
where $E_L(p)$ is the electric field that we wish to measure, and $\epsilon_p e^{j \phi_p}$ represents the
instrumental leakage of the unwanted right-circular polarization electric field, $E_R(p)$, on antenna $p$. 
Similarly, the voltage output of a right-circular polarized antenna, $q$, may be written
\begin{equation}
V_R(q)=E_R(q) + E_L(q)\epsilon_q e^{j \phi_q}, 
\end{equation}
where $\epsilon_q e^{j \phi_q}$ represents the instrumental leakage of left-circular polarization on
antenna $q$.

For a point source with zero circular polarization the correlator output on the LR cross-polarized baselinem 
which combines antenna $p$ with antenna $q$, can be written
\begin{eqnarray}
V_L(p)V_R^*(q) &=& {E_L(p)E_R^*(q)e^{2j\eta}+
  I(\epsilon_pe^{j\phi_p}+\epsilon_qe^{-j\phi_q})\over
  {(1+\epsilon_p^2)^{1/2}(1+\epsilon_q^2)^{1/2}}} \nonumber \\
               &=&{mIe^{2j\chi}e^{2j\eta}+I(\epsilon_pe^{j\phi_1}+
  \epsilon_qe^{-j\phi_2})\over {(1+\epsilon_p^2)^{1/2}(1+\epsilon_q^2)^{1/2}}},
\end{eqnarray}
where  $m$ is the linear polarization, $I$ is the total intensity, $\chi$ is the polarization angle on the
sky, and we have neglected terms of order $\epsilon^2$ and higher.  The angle $\eta$ is the azimuthal
angle of the antennas about the line of sight to the source. In the CBI and the DASI, the antennas are mounted on a rotatable deck, so that, in
addition to the usual two (alt-az) axes, the instruments can be rotated about the optical axis, thereby
varying
$\eta$ for all antennas simultaneously.  The expression given above applies for instruments such as the CBI
and the DASI in which a single rotation by $\eta$ of the deck upon which the antennas are all mounted leads to a
phase advance of $\eta$ in one polarization and a phase retardation of $\eta$ in the other polarization ---
thereby giving rise to a relative phase shift of $2\eta$ on cross-polarized baselines. 

By means of rotation of the deck, it is easy to measure the instrumental polarization on an instrument
such as the CBI or the DASI:  we simply need to observe a bright unresolved polarized source, for which $m$
and $\chi$ are known, such as 3C~279,  and measure $V_L(p)V_R^*(q)$ while
varying
$\eta$ by rotating the deck.  The complex number $V_L(p)V_R^*(q)$ then
traces out an ellipse that is closely approximated by a
circle of radius
\begin{equation}
mI/(1+\epsilon_p^2)^{1/2}(1+\epsilon_q^2)^{1/2}
\end{equation} 
centered on the point 
\begin{equation}
{I(\epsilon_pe^{j\phi_p}+\epsilon_qe^{-j\phi_q})\over
{(1+\epsilon_p^2)^{1/2}(1+\epsilon_q^2)^{1/2}}} \quad .
\end{equation}  
 
By observing a bright  source of known polarization on all baselines, it is then
possible to solve for the individual antenna leakage factors to high precision.  Both the CBI and the DASI use achromatic polarizers with
leakage factors $< 3\%$ designed by John Kovac (Kovac et~al.\ 2002), so that, after correcting for instrumental
polarization, uncertainties in polarization due to instrumental effects are
$\ll 1\%$.

The polarization of the CMB can be expressed in terms of a curl-free mode and a curl mode. 
By analogy with electromagnetic theory these are designated as the $E$-mode and the $B$-mode. 
The strongest polarized signal in the CMB is due to Thomson scattering by electrons  of the
anisotropic radiation field due to quadrupole velocity anisotropy, and this is an $E$-mode
component.  The much weaker $B$-mode components would be caused by gravitational radiation or
gravitational lensing (see, e.g., Hu 2003).

It can be shown that for the case of infinitely small antennas the sum of the LR+RL visibilities yields 
the Fourier transform of the $E$-mode, while the difference LR$-$RL
yields the Fourier transform of the
$B$-mode (Zaldarriaga 2001).  For the realistic case of finite antennas  there is also a small amount of mixing of these modes due to the
finite aperture.  This is a direct analog of the arguments given above showing that the visibility in total
intensity is the convolution of the angular spectrum with the Fourier transform of the primary beam.

\subsection{Instrumental Crosschecks}
On the CBI there are many duplicated baselines in any given orientation of the deck, and further baseline
duplication can be accomplished by deck rotation.  This is very useful in tracking down and
eliminating various sources of systematic errors (e.g., cross-talk between electronic components).  It also
provides for crosschecks of visibilities on calibrators measured on different baselines, etc.

\section{CMB Spectra from Interferometry Observations}

The CBI began CMB observations at the Chilean site in January 2000, and the DASI and the VSA began their observations shortly thereafter. 
In this section we discuss the details of the spectra revealed by these interferometric observations, and in the following section we discuss the
constraints that these place on key cosmological parameters.

\begin{figure*}[t]
\centering
\includegraphics[width=1.00\columnwidth,angle=0,clip]{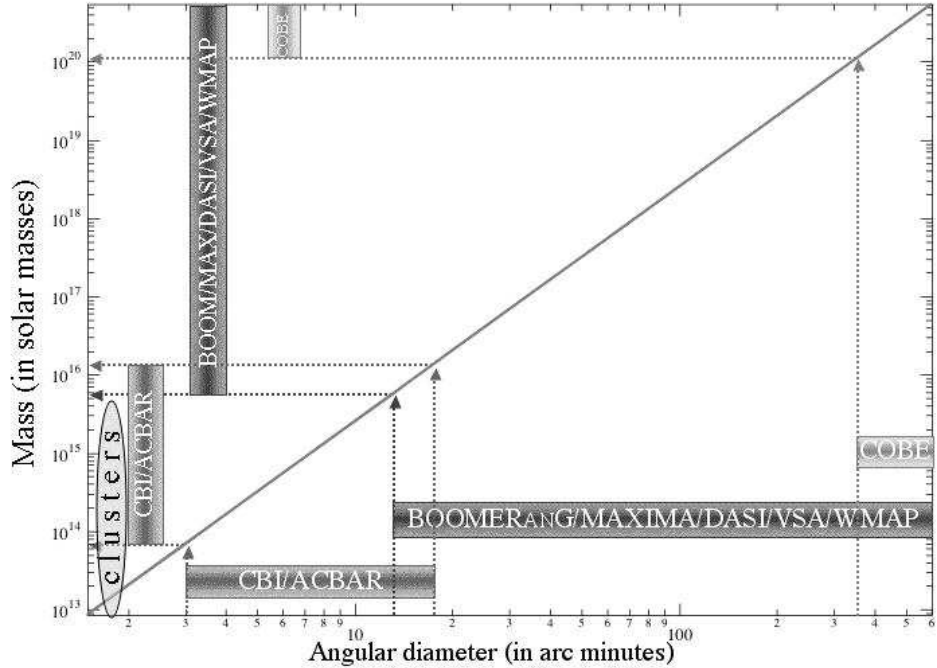}
\vskip 0pt \caption{
The mass of material within a sphere at the last-scattering surface is shown 
as a function of the angular diameter of the sphere.  Here we have assumed a 
$\Lambda$CDM model in which $\Omega_{\rm matter}\,  h^2 =0.3$ and 
$\Omega_{\rm total}=1$.  The mass scales accessed by various CMB observations 
are shown.  Only the CBI and ACBAR cover masses of galaxy clusters.
\label{fig:masses}}
\end{figure*}

\begin{figure*}[t]
\centering
\includegraphics[width=0.97\columnwidth,angle=0,clip]{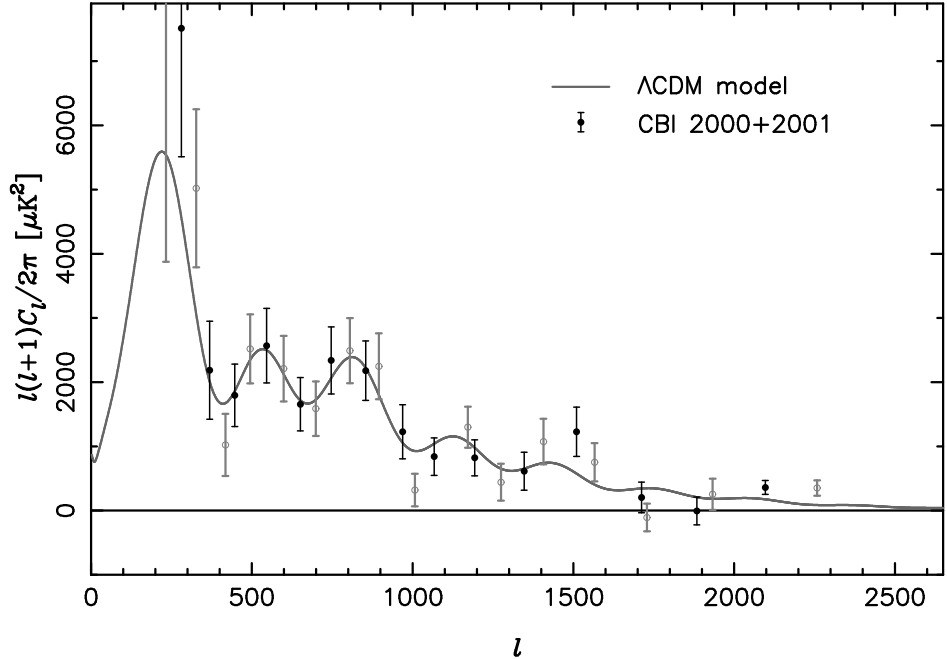}
\vskip 0pt \caption{
The CBI spectrum shown here was obtained with a combination of deep
observations of single fields and mosaic observations of 
$2^{\circ} \times 4^{\circ}$ fields.  The results for two alternate binning 
schemes are shown by the solid error bars and closed circles, and the dashed 
error bars and open circles. These are not independent points, but there is 
more information in the data than can be displayed in a single spectrum, and 
this also shows that there is nothing peculiar about the particular binning 
scheme being used. The spectrum shows four important features for the first 
time: (1) anisotropy on multipoles above $l\approx 1000$, corresponding to mass 
scales typical of clusters of galaxies; (2) the damping tail due to photon 
viscosity in the scattering region and the finite thickness of this region; 
(3) the second, third, and fourth acoustic peaks, with strong hints of the
first and fifth peaks; and (4) an excess of power in the multipole range 
$2000<l<3500$, ascribed to the Sunyaev-Zel'dovich effect in either galaxy 
clusters or the ionized regions produced by Population III stars (see text).
\label{fig:cbispectrum}}
\end{figure*}

\subsection{The CBI CMB Spectrum}
 The first results from the new generation of interferometers were published by the CBI group in January 2001. These were
the first observations of the CMB with both the sensitivity and the resolution to make images of
the mass fluctuations on scales corresponding to clusters of galaxies (Padin et~al.\ 2001; Mason et~al.\ 2003; Pearson et~al.\ 2003), and the images show
clearly, and for the first time, the seeds that gave rise to galaxy clusters (Fig.
\ref{fig:mosaics}).   The straight line in Figure
\ref{fig:masses} shows the mass within a spherical region at the epoch of last scattering as a
function of the angular diameter of the sphere, assuming a $\Lambda$CDM cosmology.  Also shown are the
angular scales and the corresponding mass scales that have been probed by key CMB experiments over the
last three years.   In addition, these were the first CMB observations to show the
reduction in spectral power at high multipole numbers due to the damping tail at
small scales caused by photon viscosity and the finite thickness of the last-scattering region.  Thus, this  major pillar of the
standard $\Lambda$CDM cosmology has been confirmed by the CBI observations. The CBI spectrum for observations from January 2000 through November 2001
is shown in Figure \ref{fig:cbispectrum}. 

An intriguing feature of the CBI spectrum is the excess of power seen at $l\approx 2500$ (Mason et~al.\ 2003).  The reality of this feature and its
possible significance are discussed below. 

At the frequency of operation of the CBI (26 -- 36 GHz), and at multipoles above $l \approx 500$, there is significant foreground
contamination due to radio galaxies and quasars.  These are dealt with in the CBI data by a constraint matrix approach, such as was
first used on interferometry data by the DASI (Leitch et~al.\ 2002a), and which effectively ``projects out'' the point sources.  The application to the
CBI is described in detail by Mason et~al.\ (2003).  Thus far it has been necessary to use the 1.4~GHz NRAO VLA Sky Survey (NVSS)
(Condon et~al.\ 1998) for identifying possible contaminating sources, because there is no higher-frequency radio survey that covers the CBI
fields.   Only
$\sim20\%$ of the NVSS sources are flat-spectrum
objects that are bright enough at the CBI frequencies to cause detectable contamination, and
a survey of NVSS sources in the CBI fields at $\sim 30$ GHz would enable us to identify these sources, thus reducing by a factor of 5 the number of
sources that must be projected out of the CBI data.  This would significantly increase the amount of CBI data that is retained, and it would
substantially reduce the uncertainties in the CBI spectrum at multipoles higher than $l\approx 1000$. It is possible to measure the flux densities of the NVSS sources in the CBI frequency range with either the Bonn 100-m telescope
or with the Green Bank Telescope.

\begin{figure*}[t]
\centering
\includegraphics[width=0.97\columnwidth,angle=0,clip]{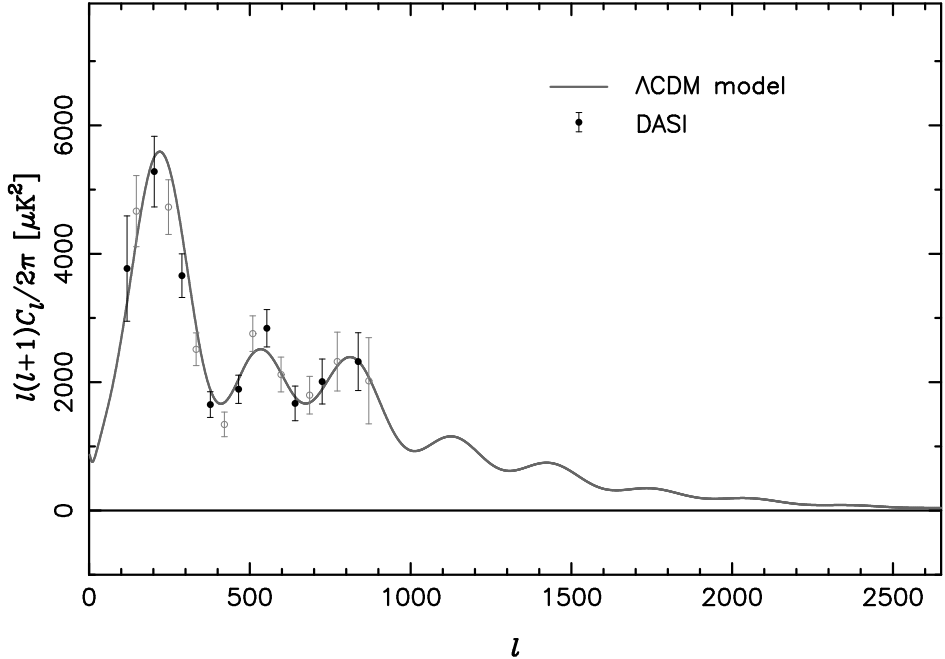}
\vskip 0pt \caption{
The DASI spectrum.  The symbols are as  for the caption to
Figure \ref{fig:cbispectrum}.
\label{fig:dasispectrum}}
\end{figure*}

\begin{figure*}[t]
\centering
\includegraphics[width=0.97\columnwidth,angle=0,clip]{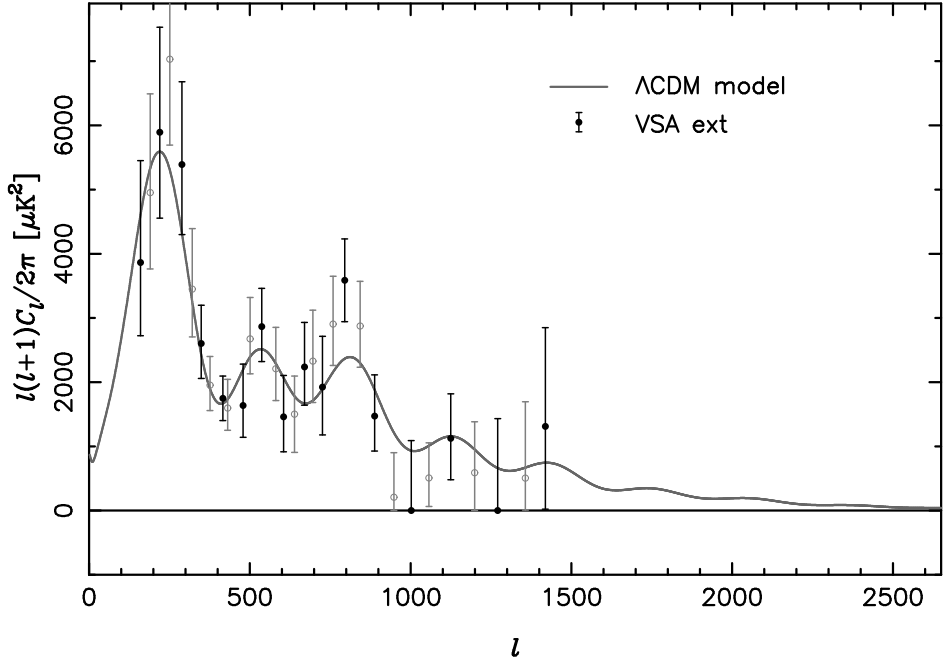}
\vskip 0pt \caption{
The VSA spectrum shown above was obtained by combining results from the first
and second sets of horns (see text).    The symbols are as  for the caption to Figure \ref{fig:cbispectrum}.
\label{fig:vsaspectrum}}
\end{figure*}

%

In 2002 the CBI was upgraded with the DASI-style achromatic polarizers, and the antennas were moved into a close-packed configuration that
concentrates the observations into the multipole range $500<l<2000$.  Since October 2002 it has been observing in this configuration.

\subsection{The DASI CMB Spectrum and the Detection of Polarized CMB}

The first DASI results were released in April 2001 (Halverson et~al.\ 2002).  The DASI
spectrum is shown in Figure \ref{fig:dasispectrum}.  We see here a clear detection of the first and second acoustic peaks, and a hint of the third peak. 
These DASI results were the first interferometry results to show the first acoustic peak, which had first been seen clearly in the TOCO observations
and then with very high signal-to-noise ratio in the BOOMERanG results released in April 2000 (de~Bernardis et~al.\ 2000).

In the second year of operation (2001) the DASI was outfitted with achromatic polarizers and carried out polarization observations that 
yielded a detection of polarized emission attributed to the CMB (Kovac 
et~al.\ 2002; Leitch et~al.\ 2002b).  Thus, another key pillar of the standard
cosmological model has been confirmed by an interferometer.  The DASI polarization results  are discussed
 in the article by Zaldarriaga (2003) in this volume.

\subsection{The VSA CMB Spectrum}

The VSA was  deployed  in 2000 with a compact array (Scott et~al.\ 2003; 
Watson et~al.\ 2003), and the following year the 
horns were replaced with larger ones and the array was extended to longer baselines, thereby permitting the VSA to
make observations of the CMB spectrum up to multipoles $l\approx 1400$.  The VSA CMB spectrum from the combination of observations with the
original and extended arrays 
 is shown in Figure \ref{fig:vsaspectrum}  (Grainge et~al.\ 2003; Scott 
et~al.\ 2003).  Here we can see clearly the first, second and third acoustic
peaks.

\begin{figure*}[t]
\centering
\includegraphics[width=1.00\columnwidth,angle=0,clip]{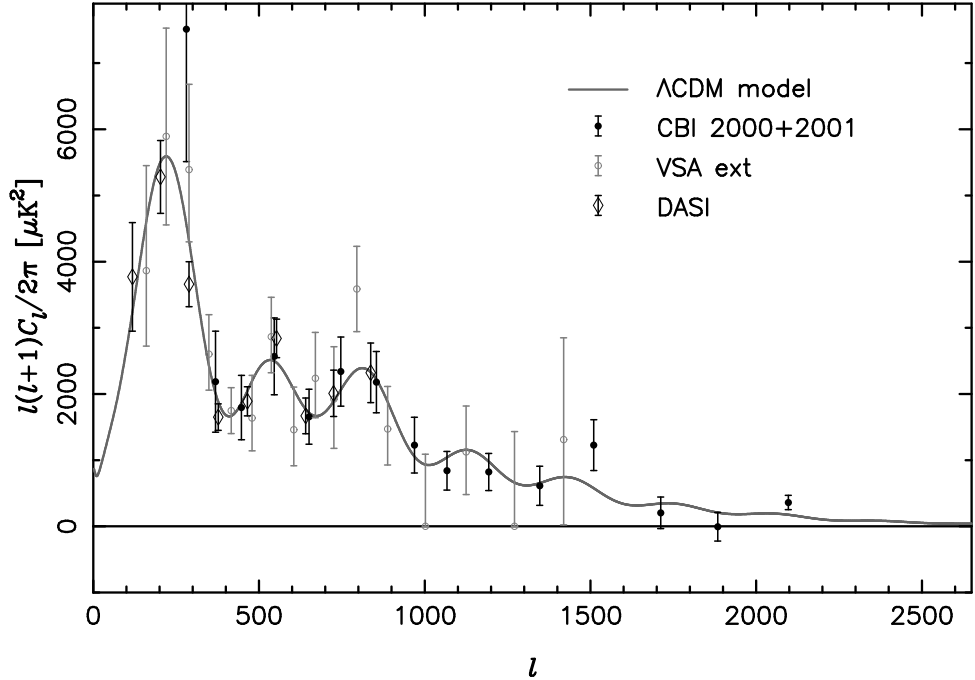}
\vskip 0pt \caption{
The combined CBI, DASI, and VSA spectra show the excellent agreement between
these three experiments in the region of overlap.  The combination of these
three spectra provides stringent constraints on key cosmological parameters
and shows a number of the key spectral features predicted by theory: a large
first acoustic peak followed by successively smaller harmonics of this peak,
with a strong damping tail.  The excess detected by the CBI at
$l\approx 2500$ was not predicted and must be confirmed (see text).
\label{fig:interferometerspectra}}
\end{figure*}

\begin{table*}[t]
\caption{Cosmological Constraints from CMB Interferometers$^{\dagger}$}
\begin{tabular}{c|cccc}
\hline \hline
     {Parameter} & {${\rm CBI}$} & {${\rm DASI}$} & {${\rm VSA}$} & {${\rm
CBI+DASI+VSA}$}\\
     \hline
\\
 $\Omega_{\rm k}$ &
$-0.10_{-0.008}^{+0.007}$&$-0.06_{-0.05}^{+0.05}$&$-0.07_{-0.08}^{+0.08}$&$-0.06_{-0.04}^{+0.04}$\\
\\
$n_{\rm s}$&
$1.03_{-0.07}^{+0.10}$&$1.03_{-0.07}^{+0.12}$&$1.07_{-0.08}^{+0.10}$&$0.98_{-0.05}^{+0.08}$\\
\\
$\Omega_{\rm cdm}\;h^2$&
$0.11_{-0.03}^{+0.05}$&$0.12_{-0.03}^{+0.04}$&$0.17_{-0.05}^{+0.06}$&$0.12_{-0.02}^{+0.03}$\\
\\
$\Omega_{\rm b} \;h^2$&
$0.040_{-0.014}^{+0.013}$&$0.023_{-0.004}^{+0.004}$&$0.034_{-0.007}^{+0.007}$&$0.024_{-0.003}^{0.004}$\\
\\
$\Omega_{\Lambda}$&
$0.62_{-0.23}^{+0.15}$&$0.58_{-0.22}^{+0.16}$&$0.43_{-0.25}^{+0.22}$&$0.58_{-0.17}^{+0.14}$\\
\\
\hline \hline
\end{tabular}
\footnotesize{$^{\dagger}$Weak prior.}
\label{tab:weakprior}
\end{table*}

\begin{table*}[b]
\caption{Cosmological Constraints from CMB Interferometers$^{\dagger}$}
\begin{tabular}{c|cccc}
\hline \hline
     {Parameter} & {${\rm CBI}$} & {${\rm DASI}$} & {${\rm VSA}$} & {${\rm
CBI+DASI+VSA}$}\\
     \hline
\\
$n_{\rm s}$&
$0.99_{-0.06}^{+0.08}$&$1.00_{-0.06}^{+0.08}$&$1.05_{-0.07}^{+0.11}$&$0.96_{-0.04}^{+0.05}$\\
\\
$\Omega_{\rm cdm}\;h^2$&
$0.15_{-0.04}^{+0.04}$&$0.14_{-0.03}^{+0.03}$&$0.19_{-0.05}^{+0.05}$&$0.13_{-0.02}^{+0.03}$\\
\\
$\Omega_{\rm b} \;h^2$&
$0.028_{-0.009}^{+0.010}$&$0.022_{-0.003}^{+0.004}$&$0.031_{-0.006}^{+0.006}$&$0.023_{-0.003}^{+0.003}$\\
\\
$\Omega_{\Lambda}$&
$0.57_{-0.29}^{+0.18}$&$0.59_{-0.26}^{+0.16}$&$0.46_{-0.28}^{+0.22}$&$0.66_{-0.20}^{+0.11}$\\
\\
\hline \hline
\end{tabular}
\footnotesize{$^{\dagger}$Flat + weak priors.}
\label{tab:weakpriorplusflat}
\end{table*}

\section{Cosmological Results}

The  CBI, DASI, and VSA CMB spectra are shown in Figure \ref{fig:interferometerspectra}
(Halverson et~al.\ 2002; Grainge et~al.\ 2003; Pearson et~al.\ 2003; Readhead et~al.\ 2003).  Here, in order to avoid confusion, we have plotted only a
single binning of the data from each instrument, but the reader should be aware that there is considerably more information than can be displayed in
this one-dimensional plot.  There is
excellent agreement between these three independent experiments in the regions of overlap.  Some of the key cosmological results from these three
interferometers are given in Table \ref{tab:weakprior}. The  ``weak prior'' assumptions used here are that
the age of the Universe is greater than 10 billion years,  $45 {\rm \; km\, s^{-1}\, Mpc^{-1}} \, < H_0 \,<\,\, 90 \,\; {\rm km\, s^{-1}\, Mpc^{-1}}$, and $\Omega_{\rm
matter}>0.1$. It can be seen
 that all three interferometers provide strong evidence for a flat geometry ($\Omega_{\rm k}\approx 0$) and a scale-invariant spectrum ($n_{\rm s}\approx 1$), both
of which are expected for inflationary universes. Given the strong evidence here for a flat geometry, we have also determined key parameters for the case of a flat
Universe, with the results shown in Table
\ref{tab:weakpriorplusflat}, where it can again be seen that the results   are in excellent agreement,
and it is clear that the matter density is only about one-third of the critical density, with the predominant matter component
being nonbaryonic.  Thus, the remainder of the energy density must be made up of something other than matter, and is here attributed to the cosmological constant.
   It is interesting to compare the CBI results, which depend almost entirely on high multipoles ($l>
600$),
and are therefore independent of the first peak,  with the DASI and VSA results, which depend entirely on low multipoles, especially on the first peak.  We
see
that the cosmological parameter values derived from both the DASI and the VSA  agree very well with those derived from the CBI.  This provides strong justification
for the assumption of a featureless, primordial density fluctuation spectrum.

  This is important because it is conceivable that the structures
observed in these angular spectra are not simply acoustic peaks, but contain significant features
that are either present in the primordial spectrum or are produced by other, as yet undetermined,
physical effects.  If this were the case there could be significant errors in the derived cosmological
parameters.  However, it is almost inconceivable that we would derive the {\it same} incorrect
cosmological parameters from two different parts of the angular spectrum.  This conclusion has been
strengthened by the recent ACBAR results (Goldstein et~al.\ 2003), which cover much of the same range of multipoles as the
CBI.

We have carried out an analysis of the CBI+{\it WMAP} observations in order to determine the additional constraints placed on cosmological parameters by the extension
of the spectrum beyond the range covered by {\it WMAP} (Bennett et~al.\ 2003; Readhead et~al.\ 2003).  The results are shown in Table \ref{tab:wmappluscbi}.  Comparison
of this table with Tables \ref{tab:weakprior} and \ref{tab:weakpriorplusflat} shows clearly the much tighter cosmological constraints from {\it WMAP}.  It can be seen
here, however, that the extension of the spectral coverage to high $l$ by the CBI yields a significant reduction in many of the parameter uncertainties.

\begin{table}[t]
\centering
\caption{Cosmological Constraints from {\it WMAP} and from {\it WMAP}+CBI}
\begin{tabular}{c|cc}
\hline \hline
$\;\;\;\;\;\;\;\;\;\;${Parameter} $\;\;\;\;\;\;\;\;\;\; $ & $\;\;\;\;\;\;\;$ {\it WMAP} $\;\;\;\;\;\;\;$ & $\;\;\;\;\;\;\;$ {\it WMAP}+CBI $\;\;\;\;\;\;\;$\\
\hline
\\
$\Omega_{\rm k}$ &
$-0.063_{-0.028}^{+0.050}$&$-0.071_{-0.023}^{+0.064}$\\
\\
$n_{\rm s}$&
$0.975_{-0.020}^{+0.032}$&$0.962_{-0.013}^{+0.022}$\\
\\
$\Omega_{\rm cdm}\;h^2$&
$0.125_{-0.0092}^{+0.015}$&$0.120_{-0.0092}^{+0.0072}$\\
\\
$\Omega_{\rm b} \;h^2$&
$0.0234_{-0.0008}^{+0.0012}$&$0.0231_{-0.0005}^{+0.0010}$\\
\\
$\Omega_{\Lambda}$&
$0.437_{-0.075}^{+0.243}$&$0.446_{-0.059}^{+0.289}$\\
\\
\hline \hline 
\end{tabular}
\label{tab:wmappluscbi}
\end{table}

\subsection{The Spectrum above $l\approx 2000$}
There is excellent agreement between all of the CMB observations in the regions of overlap, and these are well fitted by a $\Lambda$CDM model.  At
higher multipoles, however, the CBI results are not consistent with the predicted levels of the anisotropies, but show an
excess that is significant at the
$\sim 3\,
\sigma$ level (Bond et~al.\ 2003; Mason et~al.\ 2003).

In the 2001 season the CBI concentrated on mosaic observations in order to increase the resolution in $l$, whereas deep observations of a small
number of fields give the best sensitivity to observations at high $l$.  Nevertheless, the 2001 observations do contribute to this region.  The
observations from 2000 alone yield a value of $508_{-149}^{+116}\, \mu {\rm K}^2$, which is an excess of 3.1 $\sigma$ over the predicted
$\Lambda$CDM model in the multipole range $2000<l<3500$; whereas the observations from 2000 and 2001  combined yield
$360_{-95}^{+102}\, \mu {\rm K}^2$, which is significantly lower in absolute terms.  The resulting significance  relative
to the $\Lambda$CDM model has therefore dropped to 2.3 $\sigma$.  The tantalizing detection of the excess persists, therefore, albeit at lower significance, 
in the combined 2000+2001 CBI data
set (see Readhead et~al.\ 2003), and more observations are required to confirm or disprove this excess.

If real, the excess at high multipoles detected by the CBI could have significant consequences, as has been spelled out in a number of papers (e.g., 
Bond et~al.\ 2003; Oh, Cooray, \& Kamionkowski 2003).  Both of these papers ascribe the excess to the Sunyaev-Zel'dovich effect.  The
paper by Bond et~al.\ attributes the excess to the  Sunyaev-Zel'dovich effect in clusters of galaxies, while that of Oh et~al.\ attributes it to the  Sunyaev-Zel'dovich effect in hot gas resulting
from supernova explosions in the first generation of stars (Population III).  The {\it WMAP} results (Bennett et~al.\ 2003) have shown that reionization
started early, at around a redshift of 20, so it may be that both the CBI and {\it WMAP} results provide evidence for Population III stars.

\section{Conclusions}

  The results from the three CMB interferometers show excellent agreement and provide compelling evidence for
an approximately flat Universe with an approximately  scale-invariant spectrum.  The derived baryonic matter density is consistent
 with big bang
nucleosynthesis calculations, and the total matter content is only $\sim$40\% of the critical density required for a flat Universe.  Therefore, it is
clear that the major fraction of the energy density of the Universe is provided by something other than matter.  This is generally
 assumed to be  a nonzero cosmological constant.

The high-$l$ excess detected by the CBI, if real, is the most interesting result of the CMB interferometry experimental results.  It is
unexpected, and would likely be due to secondary anisotropy. Both the suggested explanations, that it is due to the
Sunyaev-Zel'dovich effect in clusters of galaxies (Bond et~al.\ 2003) or in the aftermath of  Population  III stars (Oh et~al.\ 2003), would provide
an important new window on the processes of structure formation.

\end{document}